\documentclass[conference]{IEEEtran}

\usepackage[utf8]{inputenc}
\usepackage{graphicx}
\usepackage{amsmath}
\usepackage{amsfonts}
\usepackage{multirow}
\usepackage{array}
\usepackage{cite}
\usepackage{url}
\usepackage{booktabs}
\usepackage{xcolor}
\usepackage{subcaption}
\usepackage{float}
\usepackage{enumitem}
\usepackage{tikz}
\usetikzlibrary{shapes, arrows, positioning}
\tikzstyle{block} = [rectangle, draw, text width=6em, text centered, rounded corners, minimum height=3em]
\tikzstyle{arrow} = [thick,->,>=stealth]

\UseRawInputEncoding

\title{CNN-ViT Hybrid for Pneumonia Detection: Theory and Empiric on Limited Data without Pretraining}

\author{
    \IEEEauthorblockN{Prashant Singh Basnet}
    \IEEEauthorblockA{The British College, Keele University \\
    Lalitpur, Nepal \\
    Email: prashantsbasnet@gmail.com}
    \and
    \IEEEauthorblockN{Roshan Chitrakar}
    \IEEEauthorblockA{Nepal College of Information Technology, \\
    Pokhara University \\
    Lalitpur, Nepal \\
    Email: roshanchi@ncit.edu.np}
}

\begin{document}
\maketitle

\begin{abstract}
This research explored the hybridization of CNN and ViT within a training dataset of limited size; and, introducing a distinct class imbalance. The training was made from scratch with a mere focus on theoretically and experimentally exploring the architectural strengths of the proposed hybrid model. Experiments were conducted across varied data fractions with balanced and imbalanced training datasets. Comparatively, the hybrid model, complementing the strengths of CNN and ViT, achieved the highest recall of \textit{0.9443} (\textit{50}\% data fraction in balanced) and consistency in F1 score around (\textit{0.85}) suggesting reliability in diagnosis. Additionally, the model was successful in outperforming CNN and ViT in imbalanced datasets. Despite its complex architecture, it required comparable training time to the transformers in all data fractions.
\end{abstract}


\section{INTRODUCTION}
 Pneumonia is a severe respiratory infection that inflates the air sacs in one or both lungs. It is diagnosed by observing the swellings on lung tissue\cite{americanlung2022}. The role of medical image analysis in early disease diagnosis and treatment is profound.  With the advancement in technologies, deep Learning techniques, such as Convolutional Neural Networks (CNNs) have demonstrated impressive performance, outperforming statistical methods like MSE, SSIM, PCC, MI, and SVM. However, CNN is limited in capturing local features and falls short when the consideration of global dependencies is a must, take medical image analysis \cite{9710634}. The recent breakthrough in Vision Transformer, a self-attention mechanism, has promising performance in capturing the complexities of the images\cite{dosovitskiy2020image}. However, it requires a large volume of training images to achieve satisfactory performance with a higher computational demand, which is not always possible in practical implications. Particularly in medical science, it the availability of accurate set of labeled and quality data remains a challenge, hence questioning the direct application of ViTs \cite{mauricio2023}. 
 
 Although hybridization techniques explored in contemporary studies have achieved satisfactory performance, they rely on a pretrained model and lack findings under data-constrained settings where transfer learning is inefficient. To address this gap of ViT, this research proposes a novel CNN-ViT hybrid model, trained from scratch with data constraints. The study theoretically and empirically compares the hybrid with standalone CNN and ViT in varied scenarios to demonstrate its performance when the training dataset is limited and with high complexity. Furthermore, it evaluates the performance when the models are introduced with class imbalance.

Vision Transformers, though capable of excelling in determining global dependencies, do not excel when the training dataset is limited. Meanwhile, the transformer architecture demands higher training resources, raising concerns in real-world applications because the attention mechanism demands quadratic complexity. Therefore, the study is motivated to reduce the variance of ViT by integrating CNN; with no dependency on pretrained models.

The primary objective is to explore a noble hybrid approach (CNN-VIT) that combines the salient features of CNN and ViT with a minimized generalization error when the training size is limited, and achieve a satisfactory benchmark without relying on pretrained models.

\subsection{Classical Statistical and Machine Learning Approaches}
\textbf{Mean Squared Error(MSE)}, a statistical tool with efficiency in resource utilization, had limitations in detecting structural differences, thus struggled to differentiate the visually alike images \cite{denkin2023}.

\textbf{Structural Similarity Index (SSIM)} overcame MSE’s gap by structurally differentiating the images \cite{wang2018}, yet with inaccuracies in comparing diverse color formats, as it considered image distortions to follow a Gaussian Distribution consistently \cite{denkin2023}. 

\textbf{Pearson Correlation Coefficient (PCC)} emerged as a popular choice in pattern identification by establishing a linear relationship between images \cite{neto2013}; also with an easy interpretation by providing output in the range between -1 to 1 \cite{article}. However, the inconsistency in the linearity of correlation challenges the performance in complex data, where it falls short in capturing the relationship between variables.  

\cite{pluim2003} noted the significance of \textbf{Mutual Information (MI)}  in capturing complex information of the images, such as brightness, contrast, and scale, whereas \cite{denkin2023} mentioned its advantage in capturing nonlinear correlation between images while highlighting its limitation in feature capturing and computational efficiency.

\textbf{Support Vector Machine (SVM)}, a supervised machine learning model, classifies data by drawing a hyperplane separating two classes \cite{cortes1995}, \cite{irvin2019chexpert}. It maximizes the margin, the distance between the hyperplane and the closest data points of each class. To improve its efficiency in complex tasks, techniques like ensemble in multi-class classification, feature overlapping in image annotation, and integrating visual and textual features were explored, yet with limited empirical evidence \cite{tian2018}. Despite the efforts, the inefficiency remained relevant in large or imbalanced datasets due to the direct proportion of training time to integrations in LIBSVM \cite{chang2022}. Although its performance in DICON datasets was above 99\%, it sharply declined with JPEG medical images\cite{maruyama2018}.

\subsection{Deep Learning Techniques}
\textbf{Convolutional Neural Network (CNN)} addressed the challenges such as semantic gap, feature overlapping, image segmentation accuracy, integration of visual and textual features, evaluation challenge, relevant in CBIR. Visualization techniques given by \cite{10.1007/978-3-319-10590-1_53}  bridged semantic gaps between low-level image features and high-level semantic concepts. Additionally, CNN addressed the gap of SVM in JPEG radiographical images by achieving an accuracy of 100\% in all instances; while SVM was limited to \textit{87.5}\% in CT, \textit{95.8}\% in MRI, and \textit{100}\% in X-Ray respectively \cite{maruyama2018} \cite{wu2018}. Additionally, \cite{10.1145/3065386} achieved a top 5 error rate of \textit{18.2}\% and an average of \textit{16.4}\% in 5 similar predictions on a deep convolutional network (60 million parameters) trained on \textit{1.2} million high-resolution images across 1000 classes with a \textit{1000} SoftMax output given by \textit{3} connected layers. However,\cite{9710634} noted CNN’s inefficiency in capturing long-range dependencies and relevancy overfitting with denser networks.

\textbf{Vision Transformers (ViT's)} flexible architecture captures global dependencies between input and output, hence providing an effective learning mechanism between distant sequences. Flexibility in parallelization; added advantage to achieve scalability in depth compared to the convolutional layers \cite{9710634}; the division of images into smaller patches contributes to effectively capturing global information; comprehensively contributes to outweigh CNN for detail-centric image recognition tasks. The transformer-based technique does not rely on convolution and pooling operations because it treats image classification as a sequence of processing tasks. It breaks down an image into multiple patches, which are discrete and individually meaningful. Subsequently, visual features are extracted from each of the patches by processing them through a series of attention mechanisms. These attention mechanisms capture the global and local contextual relationships, which outshine CNN in capturing complex information. \cite{dosovitskiy2020image}
explored the application of ViT, initially designed for textual tasks, in image classification. An accuracy of \textit{88.5}\% on a larger dataset (\textit{14M – 300 M} images, ImageNet); \textit{90.72}\% on ImageNet-RealL (diverse datasets); and \textit{94.55}\% on CIFAR-100 (extensive category) was achieved, signifying its impressive performance with a larger scale of training data and with images having natural and adverse disturbances \cite{mauricio2023}. At the same time, they noted lower accuracy of ViT with moderate training datasets, usually in which CNN excelled. Consequently, criticisms were raised in its application in practical resource-sensitive scenarios. \cite{unknown}

To achieve a remarkable performance in a resource-centric environment, \textbf{a light-weight parallel CNN} was used by \cite{kibria2023}  in classifying COVID-19 pneumonia with an average accuracy of \textit{99.21}\%. \cite{wu2018} combined CNN with the DNN. They used CNN as a feature extractor and provided experimental results in image forgery detection. The result had a high accuracy of\textit{ 98.2}\% on training datasets. 

\subsection{Hybridization Techniques}
\cite{tay_nodate} explored the performance of ViT and CNN in the pretrained model. CNN outperformed ViT in \textit{7} out of \textit{8} tasks, though the tasks were related to language models and natural language processing. 

\textbf{Two-Step CNN-ViT} model combined CNN (ResNet-59) with ViT(ViT-b16) sequentially, complementing the CNN for local feature extraction and ViT for global. However, it relied upon pretrained datasets to achieve an accuracy of \textit{98.97}\% for binary classification \cite{TwoStepCNNViT2024}. Likewise, \textbf{Hybrid Convolutional Vision Transformer (HyCoViT)} combined CNN and ViT for the same purpose \cite{HybridCNNSurvey2024}.

\textbf{LungMaxViT} leveraged MaxViT's multi-axis attention, based on maxvit pretrained with ImageNet 1K datasets, to classify Chest X-rays/ CT scans to detect lung diseases. The model achieved a classification accuracy of \textit{96.8}\% and F1 scores of \textit{96.7}\% on the Covid-19 dataset; and an AUC score of \textit{93.2}\% and F1 scores of \textit{70.7}\% on the Chest X-ray 14 dataset using enhancement techniques \cite{Fu2025LungMaxViT}.

\textbf{EfficientViT} achieved up to 6 times faster inference than \textbf{MobileViT}. The latter outperformed MobileNetv3 and processed with ViT global context via unfolded attention with linear complexity of \(\mathcal{O}(HW d)\) \cite{MobileViT2021}.

\textbf{Swin Transformers} demonstrated linear complexity of \(\mathcal{O}(HW d)\)  by using hierarchical, shifted window-based attention and avoiding CNN inductive bias. However, it is resource-intensive, as mentioned by \cite{SwinTransformer2023}.

At the same time, it is worth noting that the performance of a hybrid approach with no dependency on a pretrained model has not been actively researched. Therefore, the research focuses on proposing the hybrid model and training it from scratch on a limited training dataset with a theoretical and experimental study to examine its reliability.

\section{METHODOLOGY}
\subsection{Research Approach }
After reviewing the literature and identifying the strengths and gaps of each chosen model - CNN and ViT, the research was driven towards integrating their architectural design to develop a hybrid deep learning technique with the goal to combine the strengths of each and also compare its performance with standalone CNN and ViTs in identifying pneumonia on chest X-ray images. To study the architectural advancement, a comprehensive theoretical comparison is made in~\ref{sec:theoretical analysis}. To experimentally justify the theoretical findings, CNN, ViT, and the proposed Hybrid models were trained with varying sizes of the training data across multiple seeds, hence obtaining mean results. In addition to the balanced dataset, the performance of the model was evaluated and compared to the contemporary performance of CNN and ViT on an imbalanced dataset.

\subsection{Data Source and Preparation }\
A comprehensive Chest X-ray image available at \cite{irvin2019chexpert}, with \textit{224,316} radiographs of \textit{60,316} patients, was taken, each with metadata saved in a CSV file, labels \textit{1,0, and -1} indicating pneumonia, normal cases, and unidentified cases. Also, the 'Frontal/Lateral' column in the data dictionary indicated whether an image was taken from a frontal or lateral view, where the 'frontal' (posteroanterior) view was chosen. To enhance the dynamics of the source, additionally, a smaller training set of chest X-rays where the images were readily arranged in two sub-folders, viz., normal (asymptomatic) and pneumonia (symptomatic), was added \cite{mooney2018chest}. Finally, the training dataset was prepared with a total of \textit{4984} Pneumonia and \textit{4913} Normal cases after segregating \textit{8}\% of the images (combining pneumonia positive and pneumonia negative) as test data.

\subsection{Data Augmentation}
The model trained based on the above dataset can incorrectly predict the minority class and give a biased output \cite{vandenGoorbergh2022} as a distinct class imbalance is observed. The normal cases were comparatively few compared to the pneumonia images; thus, normal data were augmented and increased by about \textit{30}\%. The techniques such as rotation, width and height shift, zooming, flipping, brightness adjustment, and scaling were randomly applied programatically \cite{10.1007/978-981-10-5272-9_39}.  

\subsection{Image Preprocessing}
\begin{enumerate}
    \item ImageDataGenerator class from the Keras module was employed to resize the images to a fixed size of \textit{128x128} pixels.
    \item To enhance the convergence of the neural network, the pixel values were normalized to the \textit{range [0,1]} by dividing by \textit{255.0} using the Rescale parameter in ImageDataGenerator class.
    \item Augmentation was applied to the training dataset by rotating the images up to \textit{20} degrees, introducing positional variations with a width shift and height shift range of \textit{0.2}. To add distortions, a shear transformation in the range of \textit{0.2} was randomly applied. Also, zoom transformation up to\textit{ 20}\% and horizontal flip of +/-\textit{20} with fill mode set to ‘nearest’ was added to ensure the original image characteristics are preserved while pixel filling is considered.
    \item Batch size controls the number of images processed together \cite{radiuk2017}. Considering the binary classification, it was set to \textit{32}.
    \item Data Splitting Strategy: Almost \textit{90}\% of the images are used for training, and close to 10\% for testing. Stratified splitting is employed to ensure a proportional representation of the classes. \cite{song2023}
\end{enumerate}

\subsection{Model Architectures}
\subsubsection{\textbf{CNN Model}}
The first convolutional block has \textit{32} convolutional filters, a \textit{3x3} kernel,  ReLU activation, Batch Normalization, max pooling with a \textit{2x2} reduction to down sample feature maps \cite{he2016deep}. The second block with 64  filters is configured with same configuration. The third and fourth blocks have the same configuration, padding, but without batch normalization or max pooling to maintain feature map size.

After these layers, a Global Average Pooling layer converts the image into a single vector. Dropout (0.1) is maintained to prevent overfitting. Finally, the output is processed by a single neuron with Sigmoid activation for binary classficiation.

\begin{figure}[htbp]
\centering
\caption{A standard CNN Architecture}
\includegraphics[height=0.2\textwidth]{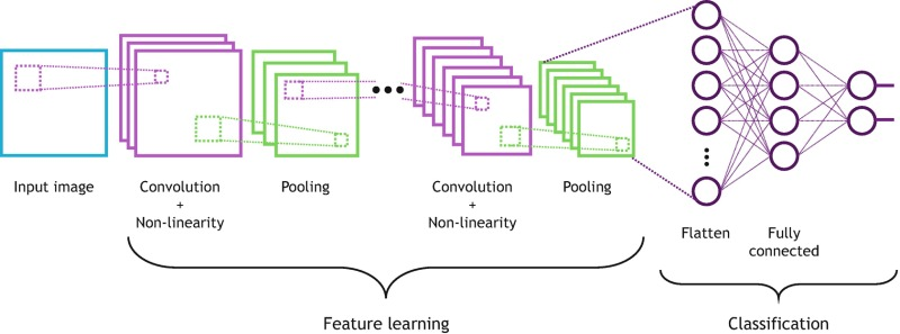}
\label{fig:cnn}
\end{figure}

\begin{figure}[htbp]
\centering
\caption{Flow diagram of the CNN model used in this research. Each component is represented in a separate box for clarity.}
\includegraphics[height=6cm]{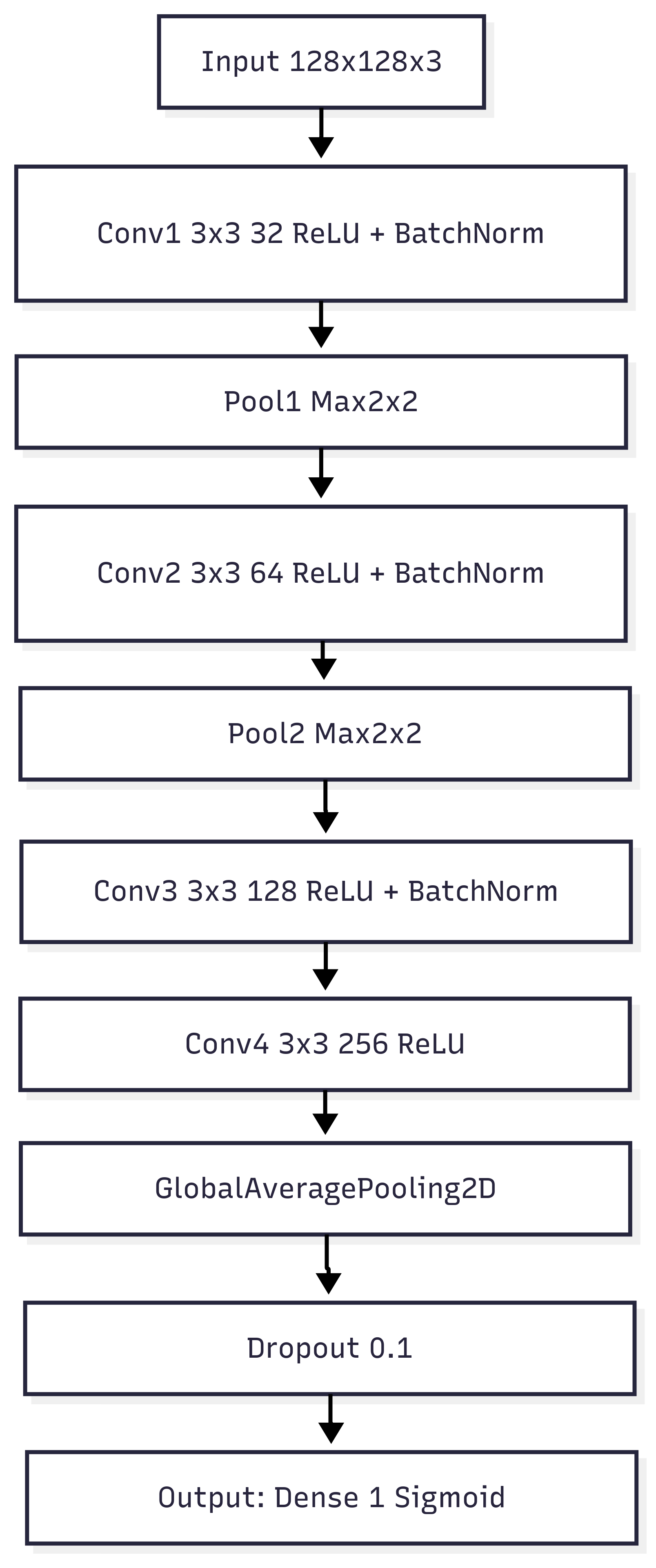}
\label{fig:cnn_archi}
\end{figure}

\subsubsection{\textbf{ViT Model}}
The input radiographic images of size \textit{128x128x3} are processed into 16x16 non-overlapping patches, resulting in \textit{64} patches arranged in an \textit{8x8} grid. To retain the spatial context, each patch is linearly embedded into a \textit{32}-dimensional vector and enriched with positional encoding via a learnable embedding layer. The model captures non-local dependencies and subtle diagnostic features as a series of \textit{4} transformer blocks, each with multi-head self-attention having \textit{2} attention heads, each of \textit{16} dimensions, that act upon it. Therefore, the total attention dimension is \textit{32}. Residual connections and layer normalization in each block stabilize optimization.

\begin{figure}[H]
    \centering
    \includegraphics[height=8cm]{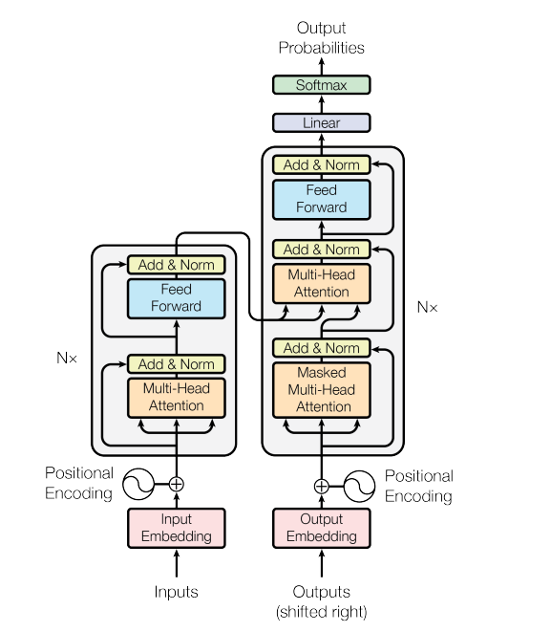}
    \caption{A standard ViT model \cite{dosovitskiy2020image}}
    \label{fig:vit_standard}
\end{figure}

\begin{figure}[H]
    \centering
    \includegraphics[height=8cm]{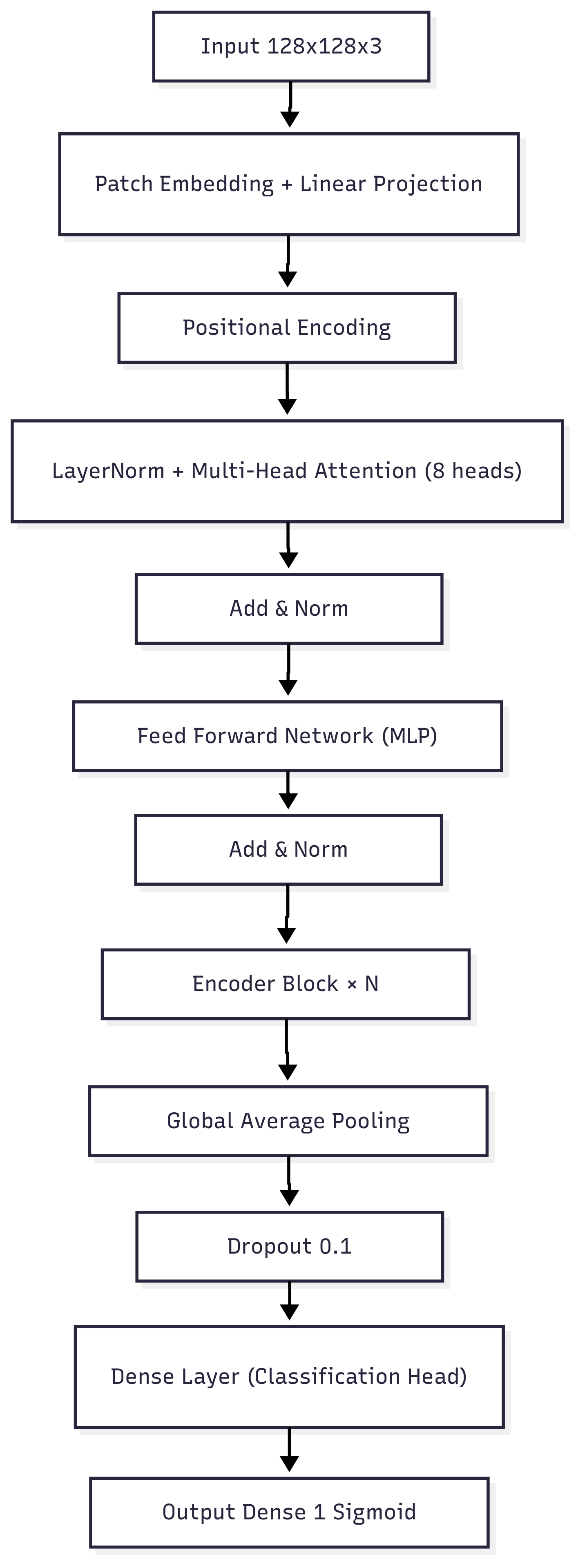}
   \caption{ViT model used in the study}
    \label{fig:vit_archi}
\end{figure}

Global average pooling, followed by a \textit{2}-layer MLP with \textit{128} \& \textit{64} neurons activated by GELU and regularized with \textit{0.3} dropout, is employed for classification. A final sigmoid layer gives the probabilistic score for pneumonia detection. \textbf{AdamW optimizer} with a learning rate of $3 \times 10^{-4}$ and smoothing \textit{0.1} is used to prevent overfitting and thus achieve stable learning.
\cite{dosovitskiy2020image}.

ViT captures subtle diagnostic indicators that the traditional convolutional approaches might not. The adaptive learning paradigm enables the model to handle variations in imaging conditions. Despite having a complex architecture, the model relies on a minimized parameter count compared to convolutional networks, making it computationally superior and faster with a faster training cycle \cite{huo2023}.

\subsubsection{\textbf{Hybrid Model (CNN + ViT)}}
The proposed hybrid model integrates CNN and ViT to leverage their complementary advantages – local feature extraction and global contextual dependencies, respectively.

Alike the standalone CNN used in the experiment, the configuration maintained in the hybrid architecture is: \textit{32, 64, 128, and 256} filters (\textit{3*3} kernels), batch normalization, ReLU activation,  same padding, and max pooling for stable learning, spatial invariance, and computational efficiency \cite{rawat2017}, \cite{zhao2024}. The input images of size \textit{128*128*3} are processed by the CNN layer and produces output image of (\textit{32*32*256}) i.e.  \( F_{\text{CNN}} \in \mathbb{R}^{32 \times 32 \times 256} \)

From the standard equation of CNN, \cite{goodfellow2016deep,oshea2015introduction},  we have
\begin{equation}
    Y_{i,j,c'} = \sum_{c=1}^{C_{\text{in}}} \sum_{m=0}^{k-1} \sum_{n=0}^{k-1} W_{m,n,c,c'} \cdot X_{i+m,j+n,c} + b_{c'}
\end{equation}
\label{eq:cnn}

where,  
$X \in \mathbb{R}^{H \times W \times C_{\text{in}}}$ is the input feature map;
$Y \in \mathbb{R}^{H' \times W' \times C_{\text{out}}}$ is the output feature map;  
$W_{m,n,c,c'}$ are the weights of the convolutional kernel; 
$b_{c'}$ is the bias term;  
$k$ is the kernel size;  
$(i,j)$ represents spatial relation; and
$C_{\text{in}}$   
$C_{\text{out}}$ representing input and output channels respectively by dividing it into , forming a \( 2 \times 2 \) grid of 4 patches.

The input images are divided \textit{16*16} non-overlapping patches (\textit{2x2 grid of 4 patches}), which are flattened into vectors of \textit{(16 * 16 * 256 = 65536)} dimensions. Then, linearly projected into a \textit{256}-dimensional embedding space to reduce their dimensionality for transformer input. 

The equation to describe the ViT model is given by:

\begin{equation}
\begin{split}
P_{i,j} &= \text{Flatten}(F_{\text{CNN}}[i \cdot 16 : (i+1) \cdot 16, j \cdot 16 : (j+1) \cdot 16]) \\
&\in \mathbb{R}^{16^2 \cdot 256}, \text{ reshaped to } X_p \in \mathbb{R}^{4 \times (16^2 \cdot 256)}
\end{split}
\label{eq:ViT's N}
\end{equation}

\textbf{Positional embeddings} preserve spatial relationships, and the resulting sequence is processed by multi-head self-attention layers with normalization and feed-forward networks \cite{vaswani_nodate},\cite{turner2024introductiontransformers}. This configuration enables the model to weigh inter-patch relationships and capture subtle anatomical details essential for clinical interpretation – the fundamentals of the medical images. 

\textbf{Multi-Head Self-Attention} is the core computational bottleneck in ViT \cite{dosovitskiy2020image}, and is given by: 
\begin{equation}
\text{head}_i = \text{softmax}\left(\frac{(X_{\text{norm}}W_{Q,i})(X_{\text{norm}}W_{K,i})^T}{\sqrt{32}}\right)(X_{\text{norm}}W_{V,i})
\label{eq:MHSA ViT}
\end{equation}

And, Feedforward with Regularization (\textit{value used in experiment = 0.5}) to prevent overfitting is given by: 
\begin{equation}
X_{\text{mlp}} = \text{Dropout}_{0.5}(\text{GELU}(X_{\text{mlp norm}}W_1 + b_1))W_2 + b_2
\end{equation}
    
\textbf{Global average pooling (GAP)}, a technique to prevent overfitting, is employed in the classification head to aggregate features, i.e., reducing high-dimensional feature maps into compact representations \cite{qiu2018}.

To prevent overfitting and enhance generalization, \textbf{a sigmoid activation layer} with output in a probabilistic score (0-1) for pneumonia presence is employed. This ensures an interpretable prediction for clinical decision-making.
 
\begin{figure}[H]
    \centering
    \includegraphics[width=0.4\textwidth]{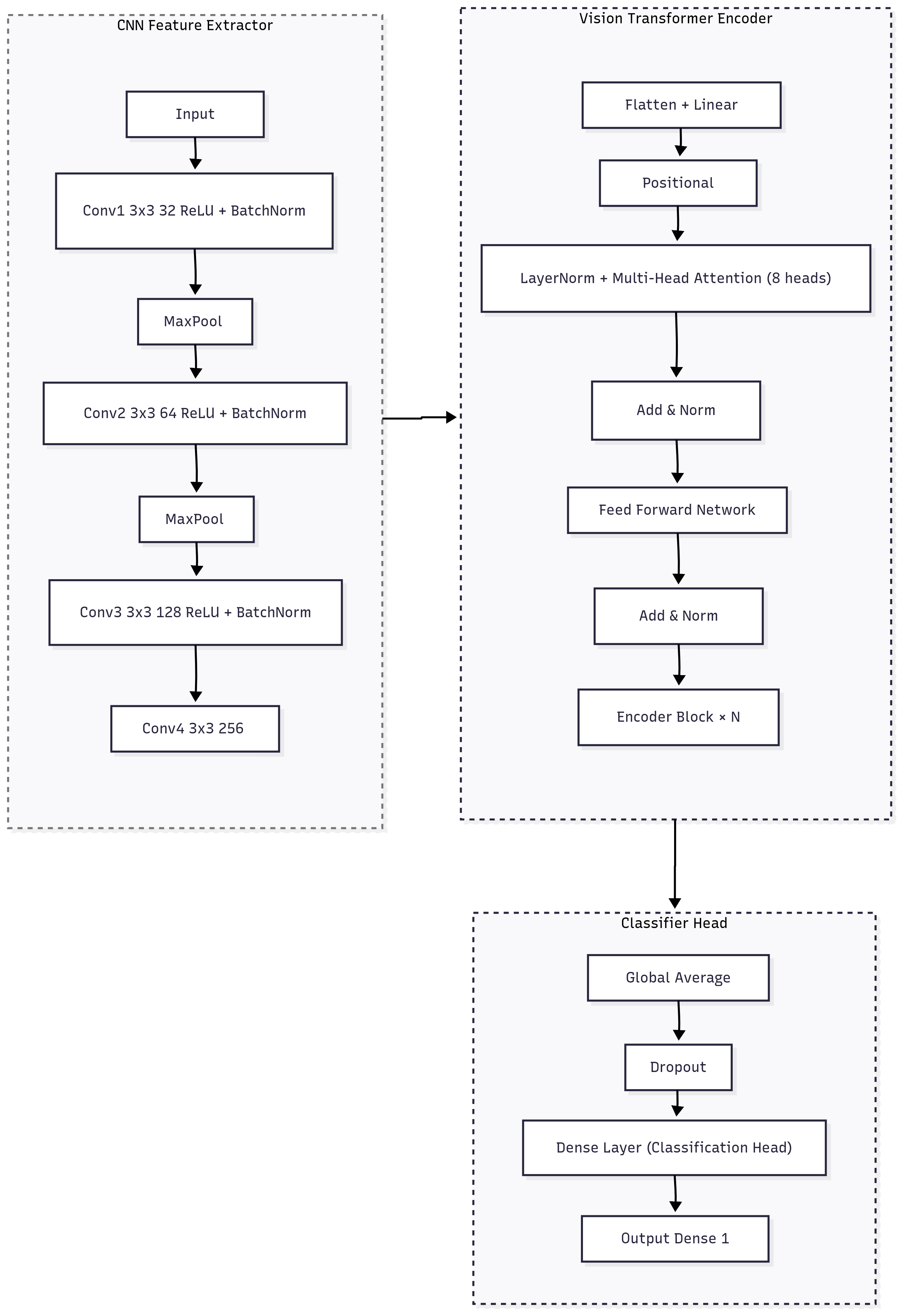}
   \caption{Proposed Hybrid Architecture with CNN for local feature extraction and ViT for global dependencies}
    \label{fig:hybrid architecture}
\end{figure}

\section{Theoretical Analysis}
\label{sec:theoretical analysis}
The equations presented above lay a theoretical foundation to compare the complexity of the model for the given dataset.

Let $H$, $W$ denote the spatial dimensions and $C_{\text{in}}$, $C_{\text{out}}$ denote the input and output channels respectively. \textit{k} is the kernel size (squared of \textit{k * k} kernels) of the convolutional filter. The computational complexity of a CNN layer is linear in pixels, given by:
\begin{equation}
    O(k^2 \cdot H \cdot W \cdot C_{\text{in}} \cdot C_{\text{out}})
\label{eq:cnn_complexity}
\end{equation}

Deriving from Equation \eqref{eq:MHSA ViT}, the complexity of ViT is quadratic per layer, given by 
\begin{equation}
    O(N^2 d) 
\label{eq:ViT's complexity}
\end{equation}

{\subsection{Challenge with ViT}}
From \eqref{eq:ViT's complexity}, ViT's capacity grows significantly with more patches. In our case, the number of images is small: \textit{4.5K} in the least training cycle. As described above, each image is split into \textit{16x16} patches, hence \textit{N=256} per image, resulting in  $256^2$, i.e.\textit{ 65,536} operations only for the attention matrix per attention head per layer as given by \eqref{eq:ViT's N}. Hence, the computational efficiency is compromised by a fixed quadratic computational cost even for limited training data.

{\subsection{Proposed Hybrid Bridges the ViT's gap by using CNN as Down Sampler}}
Generalization error bound is given by:
\begin{equation}
    \epsilon \leq \sqrt{\frac{\mathrm{VC}(\mathcal{H}) \cdot \log(1/\delta)}{N}}
\label{eq: GC}
\end{equation}
In equation \eqref{eq: GC}, VC is the \textbf{Vapnik-Chervonenkis dimension}, a theoretical measure to denote model's complexity \cite{bartlett2017nearly}. A deep neural network with a higher VC, for example, ViT model used in the research paper, memorizes noise of the small training dataset as the bound becomes loose due to smaller\textit{ N}, hence high \textit{VC}.

The minimization of generalization error in hybrids is possible with the integration of CNN layer. 

From the pure ViT model and CNN-ViT model used in the study, the number of patches are \textit{64} and \textit{4} respectively. Plugging the values in \eqref{eq:ViT's complexity}:
\begin{table}[h]
    \centering
    \caption{Sequence Length and Complexity Comparison}
    \begin{tabular}{lccc}
        \toprule
        Model             & $N$ & $N^2$ & Rel. Comp. \\
        \midrule
        Pure ViT (128$\times$128 image) & 64 & 4,096 & 256$\times$* \\
        Hybrid (32$\times$32 feat. map) & 4  & 16    & 1$\times$ \\ \\
        \bottomrule
    \end{tabular}
    \caption*{*Relative Complexity includes scaling from projection dimension and number of layers.}
    \label{tab:complexity_comparison}
\end{table}

Clearly, 
\begin{enumerate}
\item CNN drastically compressed the large, low-level\textit{ 128*128*3} image into a small, high-level \textit{32*32*256} feature map. Compared to the pure ViT, the sequential length for hybrid model has thus been reduced by \textit{16 X}, hence reducing the compression complexity by \textit{256 X}.
\item Due to reduced number of patches from \textit{64} (in ViT) to \textit{4} (in hybrid), generalization gets better. It uses CNN to reduce \textit{N} as CNN processes the image to extract local features and pass the down sampled (\textit{N}) output to the ViT. 
\item Hybrid bridges the gap of ViT by reducing VC value, thereby avoiding the overfitting of ViT while enhancing the feature extraction capabilities of CNN, derived from the equation \eqref{eq: GC}.
\end{enumerate}

\section{EXPERIMENT I : BALANCED DATASET}
\subsection{Training and Evaluation}
Computational Resource: All the models were trained in the Keras environment with a TensorFlow \textit{GPU T4x2} backend. However, computational constraints limited the ability to explore large-scale training, which could have further strengthened the findings, particularly in justifying theoretical complexity experimentally.

\subsubsection{Experimental Setup}
CNN, ViT, and Hybrid models were trained consistently with an identical preprocessing pipeline, without dependency on pretrained weights. At the same time, comparable complexity was maintained to capture fine-grained details in the medical imaging. Key components such as pooling layers, activation functions, dropout, regularization, and batch normalization were incorporated consistently across models to preserve stability and generalization while maintaining comparable complexity to capture fine-grained details in the medical imaging.

\subsubsection{Training Strategy:}
Models were trained up to 30 epochs with early stopping based on validation loss. AdamW Optimizer with a staged learning rate \textit{($3 \times 10^{-4}$ to }$6\times10^{-4}$ to $1.2\times 10^{-4}$ was employed. A batch size of 32 with stratified sampling to preserve class imbalance was used. Data augmentation, such as rotation, flipping, shifting, and zooming, was applied using TensorFlow’s \textbf{ImageDataGenerator}.

\subsubsection{Data Utilization:}
The experiment was conducted using \textit{100}\%, \textit{70}\%, and \textit{50}\% of the training data. Each experiment was repeated \textit{3} times with seeds (\textit{42,123,456}) to ensure the variability in weight initialization, data splitting, and augmentation. 

\subsection{Evaluation Metrics}
A fixed test set was taken to ensure a comprehensive performance analysis using metrics, viz., accuracy, precision, recall, f1-score, and confusion matrix. 

The definition and the significance of the metrics are as follows \cite{vakili2020performance}:

\begin{itemize}
    \item \textbf{Accuracy} provides a general measure of correctness. It can be misleading in the imbalanced datasets.
    \item \textbf{Precision} measures the proportion of true positives among all the predicted positives. It is a key evaluation metric in disease diagnosis – The higher the precision, the lower the false positives.
    \item \textbf{Recall} measures the proportion of true positives correctly identified by the model. The higher the recall, the lower the likelihood of false negatives.
    \item \textbf{F1 Score} is the harmonic mean of precision and recall. It is crucial to determine.
    \item \textbf{The Confusion Matrix} provides a detailed breakdown of the classification, thus providing a visual and quantitative analysis of model strengths and weaknesses. 
    
\end{itemize}

\textbf{Clinical Relevance}: In medical diagnosis (pneumonia detection), recall and accuracy become critical metrics. Specifically, recall provides confidence by numerically representing the accuracy of the model in not missing true cases. Meanwhile, a broader overview is given by accuracy.

\section{RESULTS AND ANALYSIS: EXPERIMENT I}

\begin{table}[htbp]
\caption{Mean Performance Comparison of CNN, Hybrid, and ViT Models (averaged over 3 seeds)}
\centering
\small
\setlength{\tabcolsep}{4pt} 
\begin{tabular}{lccccc}
\toprule
\textbf{Model} & \textbf{Data Fraction} & \textbf{Accuracy} & \textbf{Precision} & \textbf{Recall} & \textbf{F1 Score} \\
\midrule
CNN    & 100\% & \textbf{0.8391} & \textbf{0.7968} & 0.9266 & \textbf{0.8565} \\
       & 70\%  & 0.8102 & 0.7664 & 0.9116 & 0.8327 \\
       & 50\%  & 0.8139 & 0.7659 & 0.9240 & 0.8372 \\
\midrule         
ViT    & 100\% & 0.7955 & 0.7621 & 0.8824 & 0.8174 \\
       & 70\%  & 0.7886 & 0.7635 & 0.8612 & 0.8087 \\
       & 50\%  & 0.7295 & 0.7083 & 0.8143 & 0.7576 \\
\midrule  
Hybrid & 100\% & 0.8349 & 0.7866 & \textbf{0.9355} & 0.8546 \\
       & 70\%  & \textbf{0.8235} & 0.7735 & 0.9328 & \textbf{0.8457} \\
       & 50\%  & \textbf{0.8267} & \textbf{0.7729} & \textbf{0.9443} & \textbf{0.8498} \\
\bottomrule
\end{tabular}
\label{tab:performance_metrics}
\end{table}

\begin{table}[htbp]
\caption{Training Time of CNN, Hybrid, and ViT Models (averaged over 3 seeds)}
\centering
\small
\setlength{\tabcolsep}{4pt}
\begin{tabular}{lcc}
\toprule
\textbf{Model} & \textbf{Data Fraction} & \textbf{Train Time (s)} \\
\midrule
CNN    & 100\% & 588.6 \\
       & 70\%  & 452.9 \\
       & 50\%  & 344.7 \\
\midrule
ViT    & 100\% & 896.2 \\
       & 70\%  & 650.2 \\
       & 50\%  & 391.9 \\
\midrule
Hybrid & 100\% & \textbf{961.3} \\
       & 70\%  & 598.6 \\
       & 50\%  & \textbf{511.7} \\
\bottomrule
\end{tabular}
\label{tab:training_time}
\end{table}

\subsection{Observation}
\begin{enumerate}
    \item \textbf{CNN} demonstrated a steady convergence across all data fractions. Its training time was relatively lower than hybrid and performance robust, making it a computationally viable choice.
 
    \item  \textbf{ViT} demonstrated high sensitivity to the training data size. There was a acute drop in performance with smaller datasets (\textit{50}\%-\textit{70}\%). Also, significant fluctuations were observed across seeds. 

    \item \textbf{Hybrid (CNN-ViT)} demonstrated a steady convergence across all data fractions with slightly longer training times. Precision and recall were balanced effectively, and performance was steady across all the seeds.
\end{enumerate}

\subsection{Comparative Analysis}
\subsubsection{Impact of Data Size}
\begin{itemize}
    \item The accuracy of CNN and Hybrid was\> \textit{80}\% even when the dataset was reduced to \textit{50}\%.
    \item ViT's performance deteriorated significantly with smaller data fractions.
\end{itemize}

\subsubsection{Precision vs Recall }
\begin{itemize}
    \item The recall of Hybrid is greater than that of CNN. This suggests that hybrid architecture misses fewer positive cases – a crucial point to be considered while disease detection.
    \item The F1 score of the CNN was slightly greater than the hybrid’s in the full dataset, yet with slightly lower recall.
\end{itemize}

\subsubsection{Complexity vs Training Time}

Theoretically, the quadratic complexity of the ViT was reduced by hybrid. However, it is to be noted that the hybrid architecture adds extra CNN layers before transformers, which demands feature projection overhead. Also, in a real-case scenario, the performance is determined by factors such as: parallelization, training data size, batch size vs GPU memory trade-offs, and framework employed. 

Although the data fraction is limited to accurately quantify how training time \textit{}\textit{T} scales with dataset set \textit{N}, we employed a power-law relationship given by \cite{powerlaw}: 
\begin{equation}
T(N) = a \cdot N^{b}
\label{eq:powerlaw}
\end{equation}

\begin{table}[htbp]
\centering
\caption{Power Law Exponents (\(a\)) for CNN, ViT, and Hybrid Models at Different Data Fractions. 
Values indicate scaling behavior: \(a > 1\) corresponds to super-linear scaling, \(a \approx 1\) linear, and \(a < 1\) sub-linear.}
\label{tab:power_law_exponents}
\begin{tabular}{lcc}
\toprule
\textbf{Model} & \textbf{50\% to 70\%} & \textbf{70\% to 100\%} \\
\midrule
CNN    & 0.81 (sub-linear) & 0.73 (sub-linear) \\
ViT    & 1.50 (super-linear) & 0.90 (sub-linear) \\
Hybrid & 0.47 (sub-linear) & 1.33 (super-linear) \\
\bottomrule
\end{tabular}
\end{table}

\subsection{Interpretation}
ViT's training time grew faster due to quadratic complexity. The hybrid demonstrated efficient training for small to medium datasets because CNN layers lowered quadratic complexity.

\section{EXPERIMENT II: IMBALANCED DATASET}
This experiment was conducted to investigate the performance of the hybrid model in a class-imbalanced case and compare it to CNN and ViT. The architectural setup remained the same for all models. No further augmentation techniques, such as SMOTE, oversampling, or under-sampling to address the class imbalance, were taken. 

We introduced data imbalance by distributing training images as:
\begin{enumerate}
    \renewcommand{\labelenumi}{\roman{enumi})}
    \item \textit{8874} Pneumonia Diagnosed Cases, \textit{4984} Normal Cases
    \item \textit{2908} Pneumonia Diagnosed Cases, \textit{4984} Normal Cases
\end{enumerate}

\section{RESULTS AND ANALYSIS: EXPERIMENT II}
\begin{table}[h!]
\centering
\caption{Comparative Performance of CNN, ViT, and Hybrid Models on Two Imbalanced Datasets}
\label{tab:comparison_experiments_ieee}
\resizebox{\columnwidth}{!}{%
\begin{tabular}{lcccccc}
\toprule
\textbf{Model} & \textbf{Dataset} & \textbf{Time (s)} & \textbf{Accuracy} & \textbf{Precision} & \textbf{Recall} & \textbf{F1 Score} \\
\midrule
CNN     & I (8729 Pneumonia, 4884 Normal) & 2865.07  & 0.5158 & 1.00 & 0.06 & 0.1244 \\
ViT     & I                              & 2799.30 & 0.8764 & 0.9214 & 0.8903 & 0.9056 \\
Hybrid  & I                              & 2832.64  & 0.9222 & 0.9118 & 0.9778 & 0.9436 \\
\midrule
CNN     & II (2908 Pneumonia, 4884 Normal)  & 1933.97 & 0.6644 & 0.6085 & 0.9894 & 0.7535 \\
ViT     & II                             & 1918.89 & 0.9216 & 0.9624 & 0.8833 & 0.9212 \\
Hybrid  & II                             & 1959.63 & 0.9725 & 0.9972 & 0.9496 & 0.9728 \\
\bottomrule
\end{tabular}%
}
\end{table}

\subsection{Observation: Dataset II (Imbalanced: 2908 Pneumonia, 4884 Normal)}
\begin{itemize} 

\item\textbf{Hybrid}: The hybrid achieved the best accuracy (\textit{0.9725}), precision (\textit{0.9972}), very high sensitivity with a recall (\textit{0.9496}), and the best F1 score (\textit{0.9728}).

\item\textbf{ViT}: ViT achieved an accuracy (\textit{0.9216}), precision (\textit{0.9624}), high sensitivity with a recall (\textit{0.8833}), and an F1 score (\textit{0.9212}).

\item\textbf{CNN}: CNN achieved an accuracy (\textit{0.6644}), precision (\textit{0.6085}), very high sensitivity with a recall (\textit{0.9894}), and an F1 score (\textit{0.7535}).

\end{itemize}

\subsection{Observation: Dataset II (Imbalanced: 8729 Pneumonia, 4884 Normal)}
\begin{itemize} 

\item\textbf{Hybrid}: The hybrid achieved the best accuracy (\textit{0.9222}), precision (\textit{0.9118}), very high sensitivity with a recall (\textit{0.9778}), and the best F1 score (\textit{0.9436}).

\item\textbf{ViT}: ViT achieved an accuracy (\textit{0.8764}), a precision (\textit{0.9214}), a recall (\textit{0.8903}), and an F1 score \textit{(0.9056}).

\item\textbf{CNN}: CNN achieved unsatisfactory accuracy (0.5158), precision (\textit{1.0000}), very low sensitivity with a recall (\textit{0.0600}), and the lowest F1 score (0.1244).

\end{itemize}

\subsection{Discussion }
The CNN model collapsed in Dataset I, always predicting the majority class. On the contrary, ViT and Hybrid demonstrated robust performance against training data imbalance. Therefore, transformer-based architectures mitigate skewed distributions by capturing richer representations. Comparatively, hybridization of CNN and ViT provides the best performance in both scenarios, combining CNN’s strong local feature extraction with ViT's global contextual learning.
 
Despite hybrid's complex architecture, the training time difference is negligible across the models. Therefore, the hybrid model achieved the best performance without incurring any additional computational burden.

\section{CONCLUSION AND FUTURE WORK}
The study explored the proposed novel hybrid (CNN-ViT) for pneumonia detection, trained from scratch on a limited dataset, with experiments on balanced and imbalanced datasets and comparisons against standalone CNN and ViT. The theoretical and experimental findings concluded the superiority of the hybrid model in effectively combining the features of CNN and ViT – local feature extraction and global attention mechanism, respectively. Theoretically, hybrid reduced the quadratic complexity of ViT while inheriting its strong performance. It exhibited a balanced precision and recall, a higher F1 score, and accuracy in all the datasets, hence demonstrating the effectiveness of the hybrid architecture in a real-world scenario where data quality and size can be challenging to obtain.

Furthermore, it can be extended to explore the hybridization to address problems such as the information bottleneck in CNN and a rigid ViT layer demanding higher computation. It seems feasible by experimenting with an architecture that applies multiple smaller ViTs to feature maps at multiple scales within the CNN hierarchy, to reduce computational cost, which is the highest in the current hybrid technique.

\bibliographystyle{IEEEtran}
\bibliography{references}

\end{document}